# Mobility modulation effects in a double quantum well infrared photon-detector

Ting-Ting Kang, [1,2,a] Susumu Komiyama,[1] Takeji Ueda,[1,2] Shi-Wei Lin [3], and Sheng-Di Lin [3]

[1]*Department of Basic Science, The University of Tokyo, Komaba 3-8-1, Meguro-ku, Tokyo 153-8902, Japan*
[2]*CREST, Japan Science and Technology Corporation, Honcho 4-1-8, Kawaguchi-shi, Saitama 332-0012, Japan*
[3]*Department of Electronics Engineering, National Chiao Tung University, 1001 Ta-Hsueh Road, Hsinchu 300, Taiwan*

**Abstract:** An electrically isolated quantum well (QW) island can be positively charged by incoming infrared photon, because its electrons absorb photon energy via intersubband transition and acquire enough energy to escape it. This process has been used in a double QW photon-detector. Here, we present the observation of so-called negative photon-response in such detector. Its origin is clarified to be an electron mobility reduction phenomenon resulted from the photon induced charges.

[a]Author to whom correspondence should be addressed. Electronic mail: ktt219@163.com (or kang@thz.c.u-tokyo.ac.jp)

Presently, for the single-photon detection in the long-wavelength range ($\lambda > 10 \mu$m), only semiconductor quantum devices are the mature detector. These single-photon detectors are based on a charge-sensitive sensor.[1] In this scheme, the active region is polarized by photo-excitation and the induced polarization is sensed by a nearby charge-sensitive sensor. Quantum dots (QD) detectors operated in ~10mK temperature are an example,[2] in which a semiconductor QD is a photon-active element while the sensor is served by a single-electron transistor (SET).

A more easy-fabricated and custom-friendly infrared photon detector operated at 4.2 K has been developed, which is called charge-sensitive infrared phototransistors (CSIP).[3,4,5] This CSIP detector utilizes double quantum wells (DQW) structures, in which an isolated top QW island (TQWI) serves as the active region and a separated conductive channel consisting of the low QW functions as the sensor.

Fig.1 (a) illustrates the device layout of CSIP used throughout this paper. A two-terminal conductor with a constricted region of a 50 µm width and a 400 µm length is formed by wet mesa etching with around 400 nm depth. Ohmic contacts (Source, Drain and Reset pad), which penetrates both top and low QW [Fig.1 (b)], are prepared by alloying of a 200 nm thick AuGeNi layer. The Schottky contacts, i.e. isolation gate (IG), Reset gate (RG) and coupler gate (CG), are formed by depositing a metal layer 20 nm Ti/100 nm Au. By applying negative bias to both IG and RG, the top QW electron gas below them can be depleted, forming an isolated TQWI whose area is the mesa region surrounded by IG and RG. If the negative bias in RG is removed, TQWI will be connected to low QW via reset pad and not isolated. Therefore, RG works as a switch for isolating TQWI. Above the TOWI, there is CG, which is made of periodic hole arrays for the purpose of coupler, so that intersubband transition (IST) can be excited in the QW.

The photon-signal can be seen via the plot of source drain current $I_{DS}$ against IG bias voltage $V_g$, as shown in Fig.2 (a). All the measurements are done in 4.2 K using liquid helium. The 300 K blackbody radiation from room-temperature optical components is transmitted by a cold metal pipeline and used as excitation light source. The source-drain bias voltage is maintained as $V_{DS}$=10 mV throughout this paper. The signal occurs when $V_g$ decreases below $V_g= -0.42$ V and the bias in reset gate $V_{reset}= -0.7$ V, i.e. when TOWI is electrically isolated. At that time, $I_{DS}$ suddenly increases when $V_g$ is below −0.42 V. However, when $V_{reset} = 0$ V (i.e., TOWI is not isolated), no such $I_{DS}$ increase can be observed.

As already discussed in Ref. [3, 5], such $I_{DS}$ increase can be explained by the physical picture shown in Fig.3 (a). After absorbing photons, the electrons excited by IST will escape from the active region-TQWI, leaving TQWI positively charged. For CSIP, this positive charges result in an enhanced electron density $\Delta N$ and a conductivity $\sigma$ increase (denoted as "positive" signal) in low QW, due to the capacitive coupling between two QWs. This picture can be summarized as: $\Delta\sigma = (N+\Delta N)e\mu - Ne\mu = e\mu\Delta N > 0$ ( $\mu$ is the mobility of the low QW and regarded as a constant).

However, there are also some occasions that this $\Delta\sigma = e\mu\Delta N > 0$ picture is ambiguous. This phenomenon is reflected in CSIP as "negative" signal $\Delta\sigma < 0$, i.e. a conductivity decrease in the low QW due to incoming photons. In Fig.2 (b), we encounter such a CSIP show "negative" signal. The AlGaAs/GaAs DQW wafers for this device are grown by molecular-beam expitaxy on an insulative GaAs substrate. They consist of a 500nm GaAs buffer layer, a 210-nm-thick AlGaAs/AlAs superlattice barrier, a 20 nm n-doping (Si: $8\times10^{17}$ cm$^{-3}$) $Al_{0.3}Ga_{0.7}As$ electron supplier, a 20 nm $Al_{0.3}Ga_{0.7}As$ spacer, a 30 nm non-doped GaAs QW (i.e. low QW), a 80 nm composition-graded $Al_xGa_{1-x}As$ (x= 0 →0.1) barrier layer, a 2 nm $Al_{0.2}Ga_{0.8}As$ tunnel barrier, a 10 nm GaAs QW layer (i.e. top QW), a 30 nm $Al_{0.3}Ga_{0.7}As$ spacer, a 75 nm n-doping (Si: $8\times10^{17}$ cm$^{-3}$) $Al_{0.3}Ga_{0.7}As$ electron supplier, a 5 nm n-doping (Si: $8\times10^{17}$ cm$^{-3}$) GaAs cap layer. The electron density and mobility in top QW are: $N_T \sim 3.6\times10^{11}$cm$^{-2}$, $\mu_T \sim 1.9\times10^5$ cm$^2$V$^{-1}$s$^{-1}$; in low QW: $N_L \sim 4.8\times10^{11}$cm$^{-2}$, $\mu_L \sim 9.0\times10^4$ cm$^2$V$^{-1}$s$^{-1}$.

The signal occurs when $V_g$ decreases below $V_g = -0.6$ V if reset channel is depleted ($V_{reset} = -0.6$ V). At that time, $I_{DS}$ suddenly decreases when $V_g$ is below $-0.6$ V. When $V_{reset} = 0$V (i.e., TOWI is not isolated), no such $I_{DS}$ decrease can be observed. In another experiment, the radiation outside is blocked by a thick Al film which is also placed in 4.2 K. In that case (not shown), source-drain conductivity decrease around $V_g = -0.6$ V disappears. Therefore, we are convinced that such conductivity decrease is due to photon influx. In the spectrum measurement, the photo response takes a sharp maximum at about ~15.0 μm [Fig.2 (b) inset, left low].

It is well known that for DQW, during the conductivity/gate voltage scan, some kink structures, similar to positive signals in CSIP, can appear.[6,7] The mechanisms are proposed to be resistance resonance [6] or exchange instabilities related electron transfer.[7] However, such two mechanisms can not be responsible for this conductivity decrease. Because the photon-response nature of this conductivity decrease means photons should be included into the explanation.

Therefore, one first possible explanation of the negative CSIP might be to assume the opposite process to the standard CSIP mechanism [Fig. 3(a)]: that is, the low QW is somehow photon-excited and the photon-excited electrons move to TOWI, as shown in Fig. 3(b). However, this simple interpretation is not likely to be applicable for the following reasons. First, the probability of IST in the lower QW is expected to be much lower than the one in the top QW. Secondly, the probability of electrons directly transferring to the TQWI, without invoking IST, might be possible via photon-assisted tunneling (PAT) process. In PAT process, the needed photon energy should be the energy difference between ground state in low QW and 1st excited state in top QW. Due to the pulling up top QW potential by the accumulated charges in TOWI, a continuous varied photon energy during the charging process [Fig.3 (b) right] is expected. This can be readily done by adjusting photon charging period in spectrum measurement, and we will have a wavelength tunable photo detector. Unfortunately, we fail to realize such wavelength tenability in this negative CSIP [see Fig.2 (b) inset, left down].

Now, we have to reconsider the mechanism of negative CSIP. Is it possible that photon excited electrons appear in TOWI and tunnels into low QW, leaving a positive charged TQWI, but conductivity in low QW decrease? If this is true, a direct consequence is that electron mobility in low QW decrease very significantly in spite of its enhanced electron density. The picture is summarized as: $\Delta\sigma = (N+\Delta N)e(\mu-\Delta\mu)-Ne\mu \approx e\mu\Delta N - Ne\Delta\mu < 0$. This situation is some unusual, because electron mobility is usually proportional to electron density.[9] To verify this picture, we propose a "capacitive charging" scheme.

In this scheme, charging of TQWI is realized by intentionally applying some bias to the CG gate above it. In order to eliminate the photon induced charging, the experiments below were done in dark condition,[10] which was confirmed by the absence of photon signal in $I_{DS}$ Vs $V_g$ measurement. For IG, a voltage $V_g = V_1$ (here $V_1 = -0.7$ V), whose amplitude $V_1$ can only deplete the top QW below IG, is applied. For RG, a voltage $V_{reset} = V_1 + V_{pulse}$ is applied. $V_{pulse}$ is a pulse wave with frequency $2f$, pulse amplitude is $-V_1$ (See Fig.4a/b, middle). In the pulse duty time, $V_{reset} = 0$ V, TQWI is connect to low QW via reset pad. In other time, $V_{reset} = V_1$ and TQWI is always isolated. $V_{reset}$ will act as a switch on the charging/discharging of TQWI. A square wave voltage $V_{CG}$, alternating in 0 V and $V_2$ ($V_2 = -0.04$ V here), with frequency $f$ is applied to CG (See Fig.4a/b, top). $V_{CG}$ will provide the voltage for charging TQWI through the CG/TQWI capacitance.

Fig.4 (c) illustrates how one charging/discharging cycle is going on. Previous to the charging/discharging cycle, a negative voltage $V_1$ is applied to both FG and RG, so that we get an isolate TQWI. The charging/discharging cycle can be divided into 4 steps. The electron density for low QW in step *i* is denoted as $N_i$. In each step, TQWI is always isolated and TQWI is only connected to low QW in the pulse between Step IV and I, Step II and III. These steps are explained below. (I), $V_{CG} = 0$ V, TQWI is neutral; (II), $V_{CG} = V_2 < 0$. Since TQWI is floating and electrons cannot flow in/out, TQWI is still neutral. This neutral TQWI means it will not screen the electric field from CG and the negative electric field from CG will pass through top QW, reaching low QW and reducing low QW electron density. Therefore $N_{II} < N_I$; (III), $V_{CG} = V_2 < 0$. Between Step II and III, a short positive voltage pulse comes into RG, TQWI is connected to low QW during the pulse and excessive electrons flow out to low QW, leaving TQWI positively charged and screening the negative electric field from CG.[11] As a result, for Step III, only a small negative electric field from CG reaches low QW. Therefore $N_{III} > N_{II}$; (IV), $V_{CG} = 0$V, TQWI is positively charged. Due to capacitive coupling between TQWI and low QW, electron accumulates in low QW. Therefore $N_{IV} > N_{III}$. After (IV), due to a short positive voltage pulse in RG, Reset channel turn on shortly, letting electrons flowing into TQWI during pulse and neutralize TQWI. As a result, electron accumulated in low QW disappears. The situation return to (I), starting a next cycle. And we have $N_{IV} > N_I$.

From the discussion above, we have $N_{IV} > N_I \approx N_{III} > N_{II}$, which also is schematic illustrated in Fig.4 (c). If the mobility in low QW is assumed to be constant, we will have $\sigma_{IV} > \sigma_I \approx \sigma_{III} > \sigma_{II}$. This is just what we observe in positive CSIP [Fig.4 (a)]- the trivial case. However, for negative CSIP in Fig.4 (b), $\sigma_{IV} < \sigma_I \approx \sigma_{III} > \sigma_{II}$, i.e. $\sigma_{IV} > \sigma_I$ is reversed to $\sigma_{IV} < \sigma_I$. (This is figured out in Fig.4.) Since $N_{IV} > N_I$ is already known, then we reach $\mu_{IV} < \mu_I$, i.e. for low QW, its electron mobility is reduced in Step IV compared with Step I, even its density is enhanced.

In conclusion, the observed mobility reduction supports a clear physical rules governing negative CSIP. Under incoming photon illumination, the same things happens to both negative and positive CSIP. Those things are: the electrons in TQWI are excited by IST absorption, then tunneling into low QW, leaving TQWI positive charged and increasing the electron density in low QW. However, the assumption that electron mobility in low QW is roughly a constant is violated. If electron mobility in low QW decreases significantly when TQWI positively charged, even cancels out the increased conductivity by enhanced electron density, i.e. $e\mu\Delta N - Ne\Delta\mu < 0$, it leads to a conductivity decrease, not an increase. Simply speaking, under the charge-sensitive sensor mechanism, negative CSIP differs from positive CSIP in "sensor" part, not the "charge" part.

The origin behind the mobility reduction may be a complicate issue. We propose that the mobility reduction in low QW can be due to the electron wave function in low QW spreading into the grade AlGaAs barrier, which is resulted from the pulling down of barrier potential by the electric field of positively charged TQWI.[5] On the other hand, the grade AlGaAs barrier is made of AlGaAs alloy, and serious scattering events are expected, greatly reducing the electron mobility inside the barrier. At the same time, we can not rule out the possibility that other factors, like interlayer coulomb scattering,[12] also contribute to the mobility reduction.

**Acknowledgements:** This work was supported by CREST project of Japan Science and Technology Agency (JST).

**References**

1. Susumu Komiyama, IEEE J. Sel. Top. Quantum Electron. **17**, 54 (2010) and reference therein.
2. O. Astavief, S. Komiyama, T. Kutsuwa, V. Antonov, Y. Kawaguchi, and K. Hirakawa, Appl. Phys. Lett. **80**, 4250, (2002).
3. Z. An, J. C. Chen, T. Ueda, K. Hirakawa, and S. Komiyama, Appl. Phys.Lett. **86**, 172106 (2005); J. Appl. Phys. **100**, 044509 (2006).
4. Z. An, T. Ueda, S. Komiyama, and K. Hirakawa, Phys. Rev. B, **75**, 085417, (2007).
5. Z. An, T. Ueda, S. Komiyama, and K. Hirakawa, IEEE Trans. Electron Devices, **54**, 1776 (2007).
6. S. F. Fischer, G. Apetrii, U. Kunze, D. Schuh and G. Abstreiter, Phys. Rev. B **74**, 115324 (2006); Alexander Palevski, Fabio Beltram, Federico Capasso, Loren Pfeiffer, and Kenneth W. West, Phys. Rev. Lett. **65**, 1929 (1990).
7. Y. Katayama, D. C. Tsui, H. C. Manoharan, S. Parihar, and M. Shayegan, Phys. Rev. B **52**, 14817 (1995); X. Ying, S. R. Parihar, H. C. Manoharan, and M. Shayegan, Phys. Rev. B **52**, R11611 (1995); W. Pan, J. L. Reno, and J. A. Simmons, Phys. Rev. B **71**, 153307 (2005).
8. L. P. Kouwenhoven, S. Jauhar, K. McCormick, D. Dixon, P. L. McEuen, Yu. V. Nazarov, N. C. van der Vaart, and C. T. Foxon, Phys. Rev. B **50**, 2019 (1994).
9. K. Hirakawa, H. Sakaki, and J. Yoshino, Phys. Rev. Lett. **54,** 1279 (1985).
10. Actually, even we do the Fig.4 experiments under light, the results will not change very much. Because photon charging process is a slow process, which need the time >20ms [see Fig. 2(b), inset, right]. While capacitive charging process is almost an instant process in ms time scale. In Fig.4, we use a large *f* (80 Hz) value, which means each step only continues ~3 ms (<< 20 ms) and the conductance change from photon process will not have significant influence.
11. Strictly speaking, even if top QW is connected to low QW through reset pad, top QW still can not completely screen the electric field from gate. And some electric field $E_{low}$, although it is weak enough to be neglected, can reach low QW. In our case, it is estimated that $\frac{E_{low}}{E_{top}} = \frac{d_T}{d_T + d_L + s_{t,L}} \approx 0.029 << 1$, where $E_{top}(E_{low})$ is the electric field strength between gate and top QW ( top QW and low QW); $d_T = d_L = a_0/4 \approx 2.5\text{nm}$, $a_0$ is the Bohr radius in GaAs; $s_{t,L} = 80\text{nm}$ is the distance between top and low QW. More details can be found in: J. P. Eisenstein, L. N. Pfeiffer, and K. W. West, Phys. Rev. B **50**, 1760 (1994).
12. Lian Zheng and A. H. MacDonald, Phys. Rev. B **48**, 8203 (1993).

**Figure caption**

Fig.1. (a) Optical image; (b) schematic illustration of CSIP.

Fig.2. Source-drain current $I_{DS}$ against Isolation gate voltage $V_g$ of (a) positive CSIP; (b) negative CSIP. Inset: up left: magnified view of the photon-response; up right: charging up of isolated top QW in a time scale; left down in (b): excitation spectrum of negative CSIP. The conductivity variation region is highlighted by dash curves.

Fig.3. The physical picture of (a) positive (usual) CSIP; (b) the assumed direct electron transfer picture of negative CSIP. In (b) right, dot (blue) line depicts the band structure resulted from the pulling up top QW potential by the accumulated charges in TOWI.

Fig. 4, Time traces of source-drain current of low QW in the "capacitive charging" measurement: (a) positive CSIP; (b) negative CSIP. The current change between step IV and I are figured out. The time evolution of the D-S current is monitored by observing the AC component of D-S current using a four-channel digital oscilloscope (Tektronix, Inc.). The square wave in coupler gate (frequency $f$, $f$ = 80 Hz here) and pulse in reset gate (frequency $2f$, $2f$ =160 Hz here.) are both provided by a multifunction generator (NF Corporation, WAVE FACTORY). The small incline of AC current is due to the AC amplifier used in the measurement circuit. (c) Schematic illustration of one cycle of the "capacitive charging" method. A color bar indicating electron density is given in the center.

**Figure**

Fig.1.

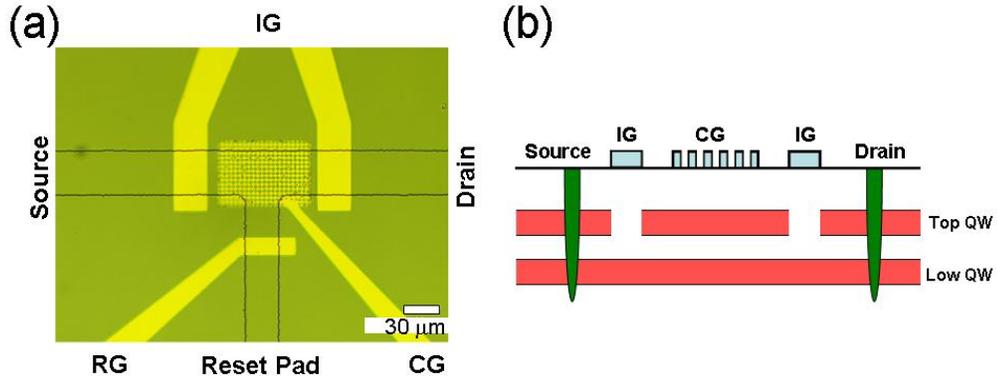

Fig.2.

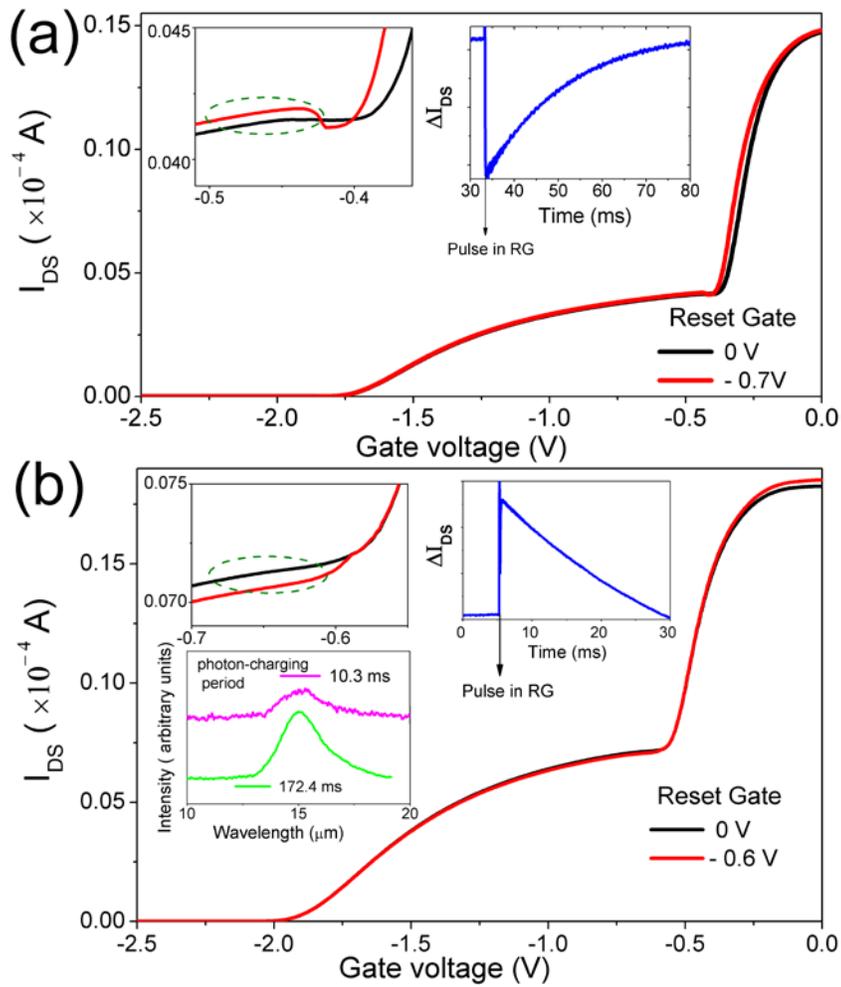

Fig.3.

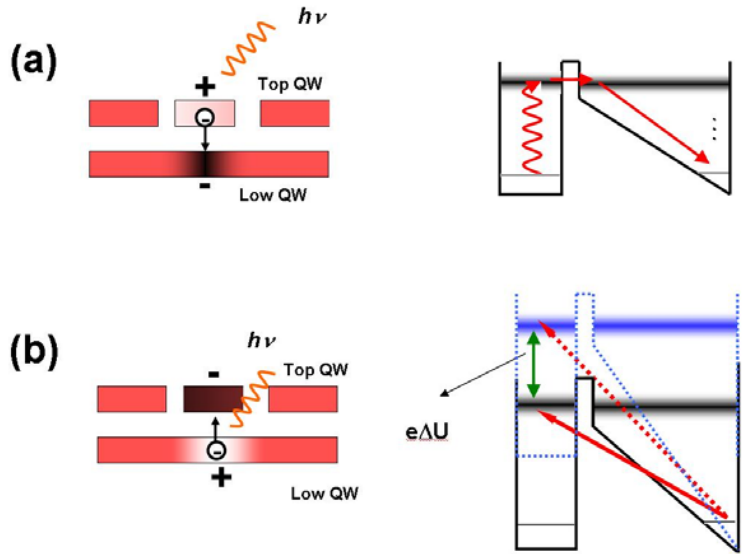

Fig. 4,

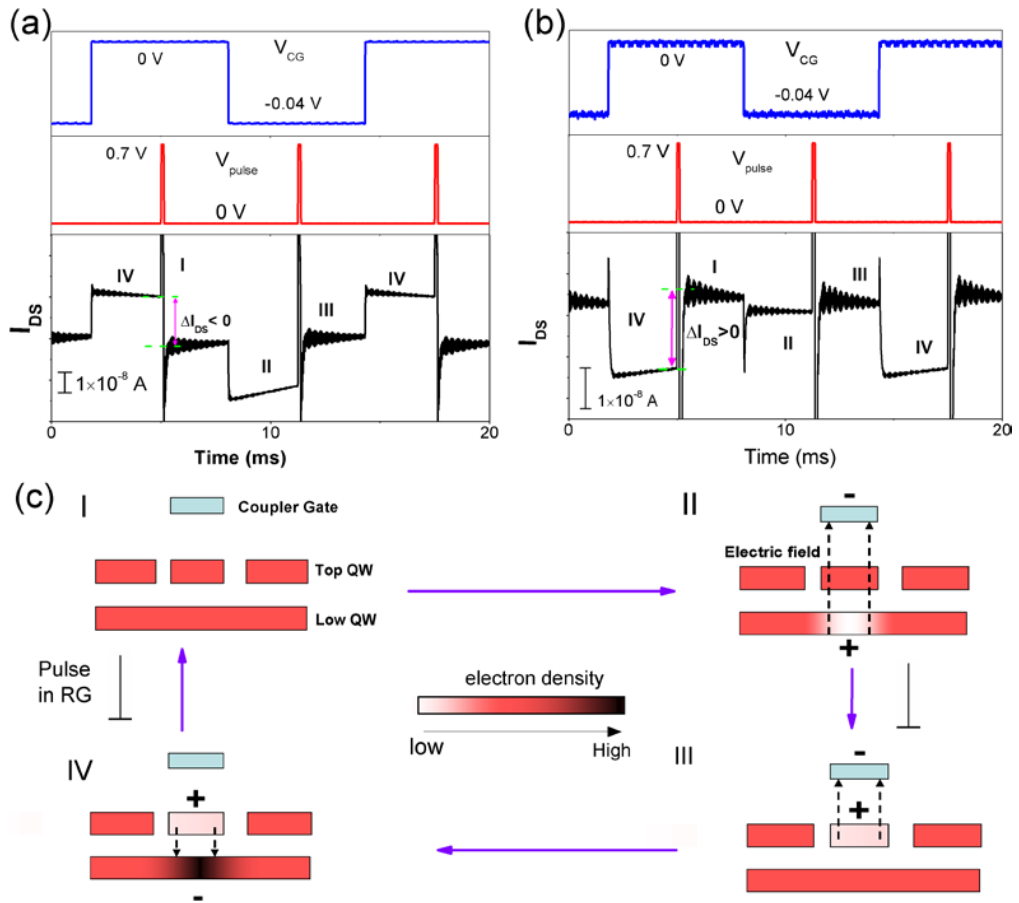